\documentclass[letterpaper,conference]{IEEEtran}
\IEEEoverridecommandlockouts
\usepackage[letterpaper, left=0.75in, right=0.75in, top=0.80in, bottom=0.79in]{geometry}

\usepackage{cite}
\usepackage{amsmath,amssymb,amsfonts}
\usepackage{graphicx}
\usepackage{textcomp}
\usepackage{xcolor}
\usepackage{marvosym}
\usepackage{multirow}
\definecolor{lightblue}{rgb}{0.2, 0.2, 0.9}

\def\BibTeX{{\rm B\kern-.05em{\sc i\kern-.025em b}\kern-.08em
    T\kern-.1667em\lower.7ex\hbox{E}\kern-.125emX}}

\begin{document}

\title{MorpheusNet: Resource efficient sleep stage classifier for embedded on-line systems
}


\author{Ali Kavoosi\textsuperscript{1 \Letter} \and 
  Morgan P. Mitchell\textsuperscript{2}\footnote{1} \and Raveen Kariyawasam\textsuperscript{3} \and John E. Fleming\textsuperscript{1} \and Penny Lewis\textsuperscript{4} \and Heidi Johansen-Berg\textsuperscript{2}  \and Hayriye Cagnan\textsuperscript{1} \and
  Timothy Denison\textsuperscript{1,3} \and
\thanks{This work was supported by the John Fell Fund of the University of Oxford, the UK Medical Research Council (MC UU 00003/3,
MC UU 00003/6) and the Royal Academy of Engineering.}\and
\thanks{\textsuperscript{1} MRC Brain Network Dynamics Unit, University of Oxford ,Oxford, UK
 }\and
\thanks{\textsuperscript{2}  
 Wellcome Centre for Integrative Neuroimaging, University of Oxford, Oxford, UK
 }\and
\thanks{\textsuperscript{3}  
Department of Engineering Sciences, University of Oxford, Oxford, UK
 }\and
\thanks{\textsuperscript{4}  
Brain Research Imaging Centre, Cardiff University, Cardiff, UK
 }\and
\thanks{\Letter :  ali.kavoosi@linacre.ox.ac.uk
 }
  }  

\maketitle

\begin{abstract}
Sleep Stage Classification (SSC) is a labor-intensive task, requiring experts to examine hours of electrophysiological recordings for manual classification. This is a limiting factor when it comes to leveraging sleep stages for therapeutic purposes. With increasing affordability and expansion of wearable devices, automating SSC may enable deployment of sleep-based therapies at scale. Deep Learning has gained increasing attention as a potential method to automate this process. Previous research has shown accuracy comparable to manual expert scores. However, previous approaches require sizable amount of memory and computational resources. This constrains the ability to classify in real time and deploy models on the edge. To address this gap, we aim to provide a model capable of predicting sleep stages in real-time, without requiring access to external computational sources (e.g., mobile phone, cloud). The algorithm is power efficient to enable use on embedded battery powered systems. Our compact sleep stage classifier can be deployed on most off-the-shelf microcontrollers (MCU) with constrained hardware settings. This is due to the memory footprint of our approach requiring significantly fewer operations. The model was tested on three publicly available data bases and achieved performance comparable to the state of the art, whilst reducing model complexity by orders of magnitude (up to 280 times smaller compared to state of the art). We further optimized the model with quantization of parameters to 8 bits with only an average drop of 0.95\% in accuracy. When implemented in firmware, the quantized model achieves a latency of 1.6 seconds on an Arm Cortex-M4 processor, allowing its use for on-line SSC-based therapies.
\end{abstract}


\section{Introduction}

\begin{figure*}[htbp]
\centerline{\includegraphics[width=1.5\columnwidth]{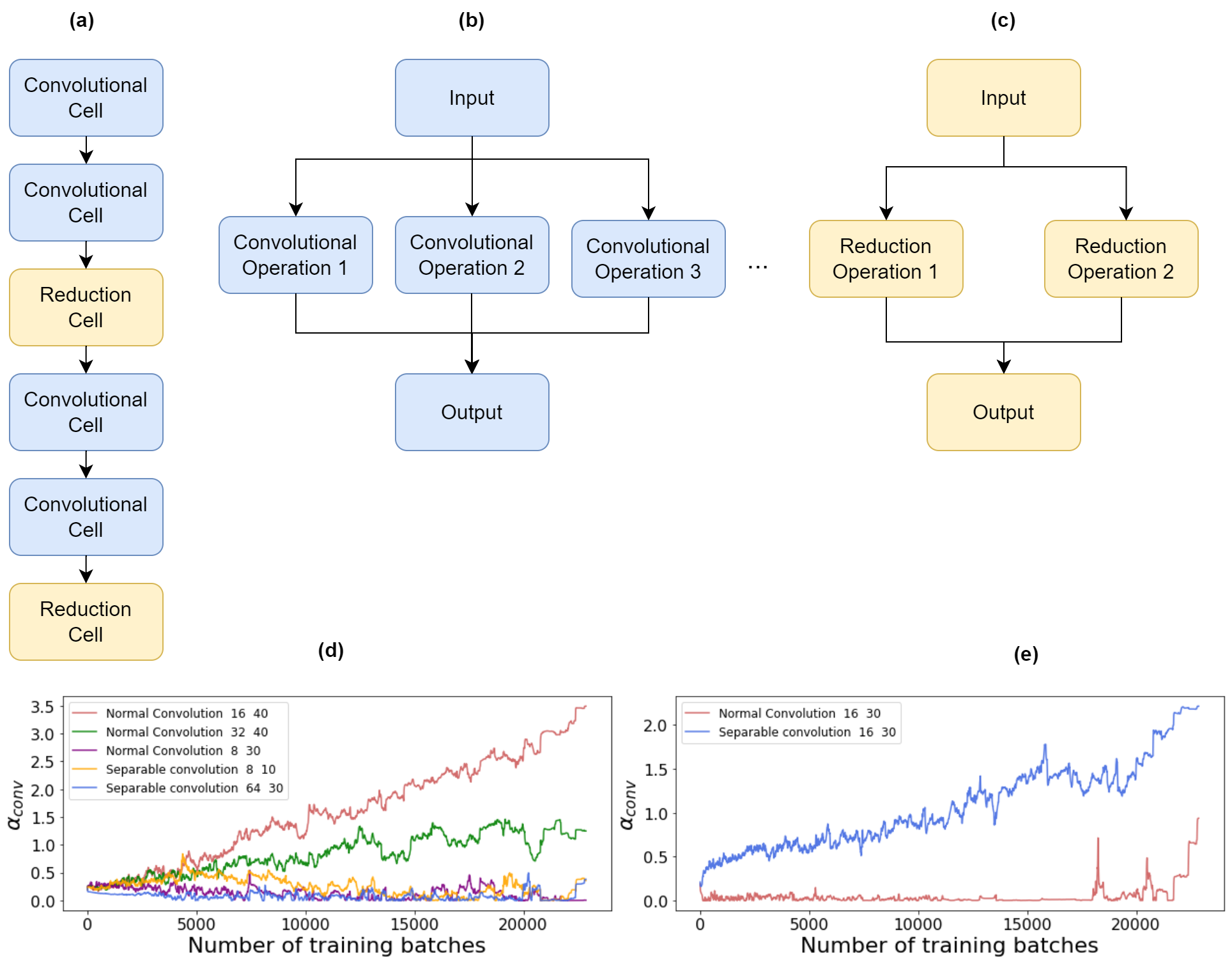}}
\caption{\textbf{(a)}: overall view of the architecture search model, light blue cells include different convolutional operations with varying filter/kernel sizes, and yellow cells are reduction cells that include either average or max pooling operations. \textbf{(b)}: architecture of a convolutional cell, with each edge having an importance factor \textbf{(c)}: architecture of a reduction cell. \textbf{(d)}: importance factor of different convolutional operations for the first convolutional cell, favoring one of the normal convolutions as the desired operation. \textbf{(e)}: Importance factor of convolutional operations in the second cell, showing the desire to choose a separable  convolution over a normal convolution with the same parameters e.g. filters and kernel size}
\label{alpha figure}
\end{figure*}
\subsection{Sleep architecture}

Sleep consists of different stages. The identification of the different stages forms an important part of different clinical applications. Diagnosis of sleep conditions, studying the sleep pattern of patients, as well as applying therapies that require to be applied at certain times during sleep require the classification of the different stages. Specifically for adaptive therapies, a further requirement will be to classify the sleep stages in real time. This will be beneficial as there is evidence that demonstrates that adaptive therapies are effective in laboratory conditions \cite{benefit}.

Sleep stages are labeled manually by experts after examining electrophysiological recordings such as Electroencephalogram (EEG) from the scalp measuring the brain activity in a non-invasive manner, Electrocardiogram (ECG) measuring the heart's rhythm and Electrooculogram (EOG) capturing the eye movements. In a clinical setting, Polysomnography (PSG) is used to record various types of electrophysiological data which are then labeled by experts using various features defined in sleep manuals, such as the American Academy of Sleep Medicine (AASM) or the R \& K standard. According to AASM, sleep can be separated into three stages: Non-Rapid Eye Movement (NREM) which consists of 3 different sub-stages in itself, Rapid Eye Movement (REM) and awake \cite{aasm}.

\subsection{Clinical Applications, Motivation for Requirements}
Targeted memory reactivation (TMR) is an experimental approach which involves pairing a memory or an action (e.g. button presses), with sensory cues (e.g., sound) during a training session. Same sensory cues are then “replayed” during sleep to achieve targeted memory reactivation. Recent studies suggest that replaying previously paired auditory cues during sleep spindles, which are key features of NREM sleep, is optimal for memory consolidation \cite{declarative}\cite{signatures}\cite{timing}, introducing a novel treatment approach for stroke rehabilitation to promote motor learning.   

TMR has also been explored in the context of mood and psychiatric disorders, based on studies highlighting a reduction in emotional arousal due to embarrassing memories following TMR during REM sleep \cite{tmrRem}.

\subsection{Requirements 2: Scalable, at-home solution}
TMR requires real-time classification of sleep stages and to-date has been carried out in sleep laboratories. This is a bottleneck for deploying such therapies at scale.

To address scalability and at-home deployment, there is a need to automate SSC and do so in real-time. There are different commercial products that perform SSC in an at-home environment \cite{dreem}. However, these products are somewhat restrictive in terms of access to raw data and classification labels in real-time. As highlighted in Section B, predicted sleep stages should be accessible to implement TMR during specific sleep stages.


\subsection{Automated SSC in energy constrained environments}

Deep Learning (DL) has been explored to automate SSC \cite{xsleep}\cite{eegnet}. Different DL architectures, such as Convolutional Neural Networks (CNNs) \cite{cnn}, Recurrent Neural Networks (RNNs) \cite{rnn} and others have been experimented with in research. The DL-based models have achieved state of the art in terms of performance. These automated approaches have reached human level performance \cite{DOD}.

As the DL models require sizable amounts of data to learn from, recent introduction of publicly available datasets have enabled accurate DL models. However, continuing research in automating SSC has shown that previous models may have been more complex than required. Recent models with smaller size and lower complexity like tinySleepnet \cite{tinysleep} have shown similar or even better performance compared to state of the art.

DL models can be used for a scalable solution to perform SSC in real-time. However, most of the models outlined to date, require sophisticated pieces of hardware\cite{compcost} and transmission of data to cloud servers or mobile devices, which introduces cybersecurity threats. Cybersecurity threats have been addressed in Food and Drug Authority (FDA) approved medical devices \cite{fda}. Therefore, a compact system that can carry out inference locally would not only save cost and consume less energy, but also improve system security. To meet these needs, we introduce MorphuesNet \cite{morph}, a highly optimized and compact model for embedded online SSC to fill this gap. Table \ref{tab:table_label_2} summarizes the requirements for MorphuesNet which are addressed in this paper.

\begin{table}[htbp]
  \centering
  \caption{Summary of requirements}
  \label{tab:table_label_2}
  \begin{tabular}{|c|c|c|}
    \hline
    \textbf{Type} & \textbf{Description} & \textbf{Aim} \\
    \hline
    \multirow{3}{*}{Real-time} & The ability to   & latency less than \\
    \multirow{3}{*}{} & classify sleep stages & 10 seconds for\\
    \multirow{3}{*}{} & in real-time & 30 second sleep epochs 
    \\
    \hline
    \multirow{3}{*}{Size} & Model should be & memory footprint to be \\
    \multirow{3}{*}{} & small enough to be & less than \\
    \multirow{3}{*}{} &  deployed on generic MCUs &  100KB\\
    \hline
    \multirow{3}{*}{Energy} &  The device needs to &  Battery to \\
    \multirow{3}{*}{} & stay operational & last at least   \\
     \multirow{3}{*}{} & long enough & 8 hours\\
    \hline
    \multirow{3}{*}{Performance} &  Whilst being small in size, & comparable sensitivity\\
    \multirow{3}{*}{} & it should maintain  & and specificity\\
    \multirow{3}{*}{} & a reasonable performance & to state of the art\\
    \hline
  \end{tabular}
\end{table}

\begin{figure*}[htbp]
\centerline{\includegraphics[width=1.5\columnwidth]{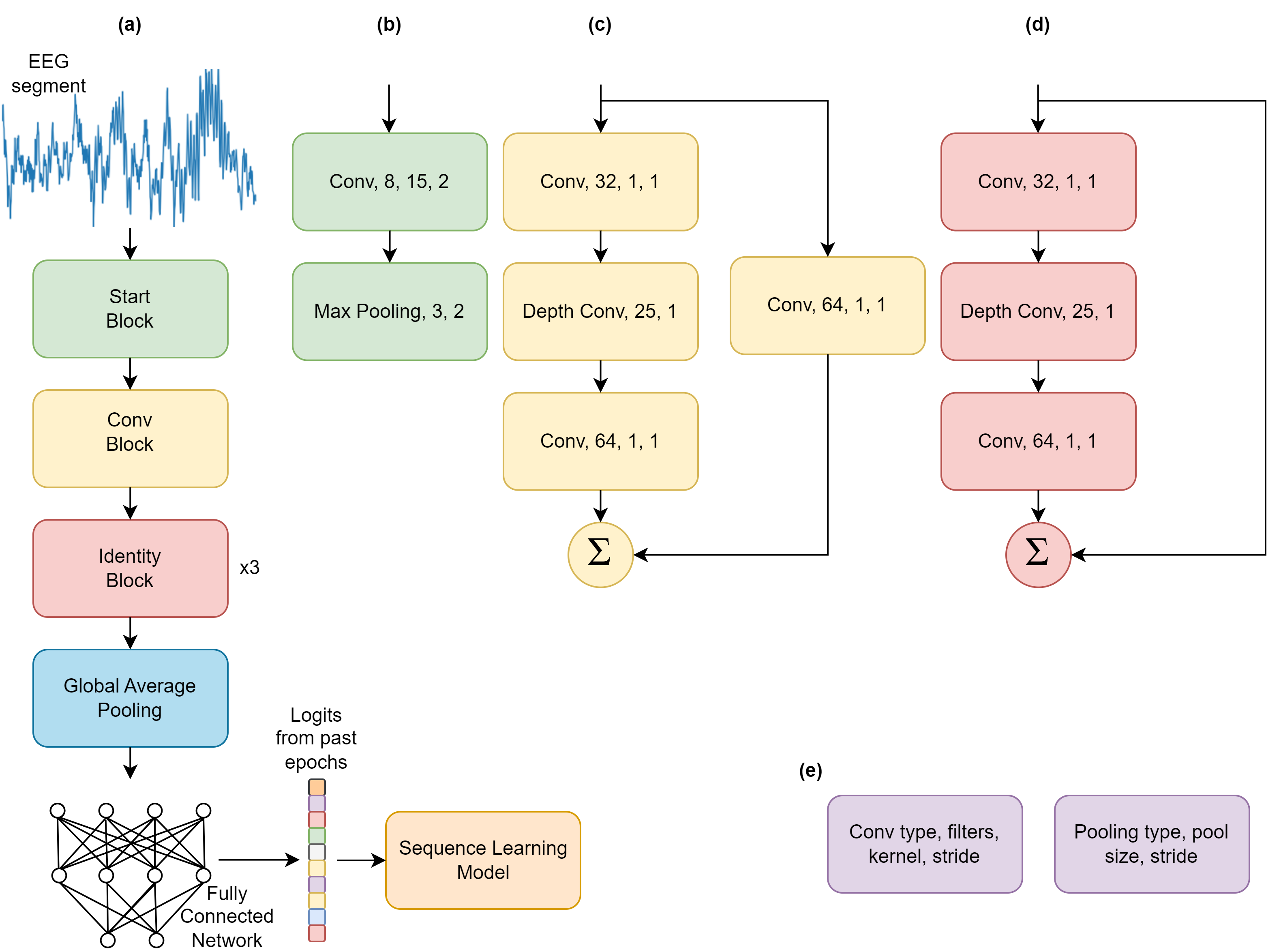}}

\caption{\textbf{(a)}: Architecture of MorphuesNet. \textbf{(b)}: Start block, which includes a normal convolution and a max pooling operation. \textbf{(c)}: Conv block, which carries out a depth-wise separable convolution and sums it with a residual connection from the input that goes through a point wise convolution. \textbf{(d)}: Identity block, which passes the input tensor through a depth-wise separable convolution and sums it with a residual connection from the input. \textbf{(e)}: Convolutional and reduction operation notations used in the figure. Each convolutional block also includes a batch normalization layer before activation.}
\label{fig2}
\end{figure*}

\section{Design principles}
We experiment with a architecture search approach. The findings are then used to design a compact network.  

\subsection{Architecture search}
We first experimented with light weight Neural Architecture Search (NAS) approaches. The search space was constrained to include two types of operations, depth-wise separable convolutions and normal convolutions for operations and max and average pooling for reduction. We deploy a Differentiable Architecture Search (DARTS) approach. However this is done by making the optimization of architecture parameters and model parameters as a single optimization problem, rather than the bi-level optimization approach used in the original paper \cite{darts}. The search is then further constrained to only have one node per layer. With this simplification, the optimization is learning a macro-architecture in a constrained search space as the architecture parameters are learned for the entire network and not for repetitive cells. To confine the model to choose between options that require less computation, the choice of kernel order and number of filters in the convolutional operations was also kept low. The maximum number of filters was fixed at 64 and the maximum kernel order was fixed at 32. The final architecture consists of four convolutional layers, and two reduction layers. This is outlined in Fig.~\ref{alpha figure}. The choice of not learning sequences was made to further reduce the computational complexity. This means that the NAS will search only for feature extraction models.

With this approach, the edges, each of which represents an operation, are allocated an importance factor denoted as $\alpha$. In this case, the optimization of the network learns both the importance of each edge and the parameters of the model denoted as $\theta$. 

Let $X$ be in the input to a cell, consisting of operations $o\textsuperscript{1}$, $o\textsuperscript{2}$ and so on, where $o\textsuperscript{i}$ is the $i\textsuperscript{th}$ operation connecting the input to the output of that cell. The function of this cell would be.
$
Y = \{\alpha\textsuperscript{1}o\textsuperscript{1}(X), \alpha\textsuperscript{2}o\textsuperscript{2}(X), 
\alpha\textsuperscript{3}o\textsuperscript{3}(X), ...\}
$, where $Y$ is the output of the cell. Simplifying this will yield:

\begin{equation}
Y = \pmb{\alpha} \pmb{o} (X)
\label{cell_func_simp}
\end{equation}

The parameters are updated using gradient decent:
\begin{equation}
\alpha\textsuperscript{*} = \alpha - \eta \frac{\partial L(\theta , \alpha)}{\partial \alpha}\label{eq1}
\end{equation}

\begin{equation}
\theta\textsuperscript{*} = \theta - \eta \frac{\partial L(\theta , \alpha)}{\partial \theta}\label{eq2}
\end{equation}

where $\theta\textsuperscript{*}$ and $\alpha\textsuperscript{*}$ are the next set of model and architecture parameters to be used, $L(\theta , \alpha)$ is the loss calculated by the model and $\eta$ is the learning rate. The choice of architecture is finalized by finding the operation with the highest $\alpha$ value between each two nodes.

Here, we observe that in the first convolutional cell, the architecture search allocates higher importance to features extracted with normal convolutions with bigger kernels. We hypothesize that this is to extract spatial features in the first layer, as there is no channel wise information in the raw input at the first convolutional operation. In this case, a normal convolutional filter can extract more useful features. 
As the goal is to upload the model on a restricted hardware platform, the possible operations were kept low in terms of computational power required. Within the domain of computationally efficient operations, the architecture search favors separable convolutions over normal convolutions. Another observation was that the model prefers max pooling in the earlier stages of the architecture and average pooling in the later stages. Fig.~\ref{alpha figure} shows $\alpha$ for two different convolutional operations in one of the cells used in the architecture search. 

We used the findings from the Neural Architecture Search experiment to design a shallow residual network  \cite{resnet}, based on depth-wise separable convolutions \cite{dsc}. To address AASM guidelines regarding transitions between states, we added a sequence learning part to the CNN's learned features. To further reduce the number of operations and parameters, the sequence learning was applied to the output layer of the CNN, which denotes the probability of different states learned separately. This part includes a Long Short-Term Memory (LSTM) based layer of 32 nodes \cite{lstm}, a Dense layer with 32 nodes with Rectified Linear Unit (ReLU) activation function\cite{relu}, where dropout was applied between the layers with a 20 \% dropout rate. 
Fig.~\ref{fig2} shows the final model architecture, referred to as MorpheusNet from hereon. The number of filters and kernel sizes were also derived from the architecture search results.
\subsection{Training}
We divide the training in 2 processes, learning feature extraction and sequence learning. Feature learning is based on the CNN. The CNN was optimized using the Adam optimizer. The learning rate was set at 0.001 and $\beta_1$ was set to 0.9 and $\beta_2$ to 0.999. The data was shuffled and divided into mini-batches of 128 samples and the model was trained for 10 epochs. The training process uses categorical cross entropy loss to update parameters. First, the model was trained on the training set. After training, the model with the best performance on the validation set was selected. The metrics were calculated on the held-out test set. 
The input to the sequence learning model consists of concatenating the output of the CNN from 12 consecutive epochs. For training the sequence learner, The same training and validation set are used. This training process has the same Adam optimizer parameters as the CNN. However, the training set is divided into mini batches of 32 samples and learning rate is set at 0.0001. The metrics reported are from the test set.

\subsection{8-bit Quantization: MorpheusNetQ}

To further simplify the model, we also experimented quantizing the model parameters. Quantization was only applied on the CNN. The quantized CNN was fine tuned to the training data for 5 epochs. The sequence learner was then fine-tuned to the quantized CNN as well. This was done by using mini-batches of 32 samples and an Adam optimizer (learning rate was 0.001) for 5 epochs. This model will be referred to as MorphuesNetQ.

\section{Dataset and data preparation}

We test the model on three publicly available datasets, Sleep-EDF \cite{sleepedf}, Dreem Open Dataset (DOD) \cite{DOD} and Physionet 2018 challenge \cite{physio2018}. 
Sleep-EDF database has two subsets, Sleep Cassette (SC) and Sleep Telemetry (ST). SC was recorded from 78 participants. The EEG channel FPz-Cz was used for classification. The dataset was expanded in 2018 from it's original 20 subjects. However, for comparison purposes, we also use the 20 subjects subset as well. For fairness in comparison, we use the same subsets for k-fold cross validation used in \cite{xsleep}. In the k-fold cross validation experiment, k was set to 20 for Sleep-EDF 20 and 10 for Sleep-EDF 78. Sleep-EDF was annotated according to the R\&K manual.
R\&K standard divides sleep stages into 6 stages (N4,N3,N2,N1,REM,WAKE) \cite{rnk}, as opposed to AASM's 5 stages \cite{aasm}. To compare the performance to other solutions, the N4 and N3 stages are merged together. Physionet-2018 database has recordings from 994 participants. we re-sample the EEG to 100 Hz, and use EEG channel C3-A2, and put k equal to 5 for cross validation. The DOD has recordings from healthy subjects (DOD-H) and patients with pathological sleep (DOD-O). We used the DOD-H subset using the EEG channel F3-M2. This subset has 25 participants and we used a leave-one-out (LOO) approach for evaluation. The EEG was re-sampled to 100Hz for this database as well. Table \ref{tab:table_label} shows the size, setup and signal of interest used for benchmarking MorphuesNet. 

\begin{table}[htbp]
  \centering
  \caption{Summary of different databases}
  \label{tab:table_label}
  \begin{tabular}{|c|c|c|c|}
    \hline
    \textbf{Databse} & \textbf{Size} & \textbf{EEG channel} & \textbf{Experimental setup} \\
    \hline
    Sleep-EDF 20 & 20 & Fpz-Cz & 20-fold CV \\
    Sleep-EDF 78 & 78 & Fpz-Cz & 10-fold CV \\
    DOD-H & 25 & F3-M2 & LOO \\
    Physionet-2018 & 994 & C3-A2 & 5-fold CV \\
    \hline
  \end{tabular}
\end{table}

\begin{table*}[h]
\centering
\caption{Summary of performance and model size of different models on different databases. }
\label{tab:mytable}
\begin{tabular}{|c|c|c|c|c|c|c|c|}
\hline
\multicolumn{1}{|c|}{Dataset} & \multicolumn{1}{c|}{Model} &
\multicolumn{1}{|c|}{Parameters(M)} & 
\multicolumn{1}{|c|}{Accuracy} & \multicolumn{1}{c|}{MF1} &
\multicolumn{1}{|c|}{Sensitivity} & \multicolumn{1}{c|}{Specificity}  \\ 
\hline
\multirow{6}{*}{Sleep-EDF 20 ($\pm$ 30 min)} & \textcolor{lightblue}{MorpheusNet} & \textcolor{lightblue}{0.02}  & \textcolor{lightblue}{84.2} & \textcolor{lightblue}{76.3} & \textcolor{lightblue}{76.9} & \textcolor{lightblue}{95.6}\\
\multirow{4}{*}{} & \textcolor{lightblue}{MorpheusNetQ} & \textcolor{lightblue}{0.02}  & \textcolor{lightblue}{84.7} & \textcolor{lightblue}{76.7} & \textcolor{lightblue}{77.1} & \textcolor{lightblue}{95.8}\\
\multirow{4}{*}{} & XSleepNet2 \cite{xsleep} & 5.6  & 86.3 & 80.6 & 80.2 & 96.4\\
\multirow{4}{*}{} & TinySleepNet\textsuperscript{\textdagger} \cite{tinysleep} & 1.3 & 85.4 & 80.5 & - & -\\
\multirow{4}{*}{} & SeqSleepNet \cite{seqsleep} & 0.2  & 85.2 & 78.4 & - & -\\
\multirow{4}{*}{} & DeepSleepNet-Lite \cite{litesleep} & 0.6  & 84.0 & 78.0 & - & -\\
\hline

\multirow{5}{*}{Sleep-EDF 20 (in-bed)} & \textcolor{lightblue}{MorpheusNet} & \textcolor{lightblue}{0.02}  & \textcolor{lightblue}{81.4} & \textcolor{lightblue}{72.0} & \textcolor{lightblue}{73.1} & \textcolor{lightblue}{94.6}\\
\multirow{5}{*}{} & \textcolor{lightblue}{MorpheusNetQ} & \textcolor{lightblue}{0.02}  & \textcolor{lightblue}{82.1} & \textcolor{lightblue}{73.9} & \textcolor{lightblue}{74.7} & \textcolor{lightblue}{95.0}\\
\multirow{5}{*}{} & XSleepNet2 \cite{xsleep} & 5.6  & 83.9 & 78.7 & 78.6 & 95.2\\
\multirow{5}{*}{} & FT-XSleepNet\textsuperscript{\textdagger} \cite{xsleep} & 5.6  & 85.7 & 80.8 & 80.6 & 96.0\\
\multirow{5}{*}{} & DeepSleepNet \cite{xsleep}\cite{eegnet} & 20  & 80.8 & 74.2 & - & -\\
\hline

\multirow{4}{*}{Sleep-EDF 78 ($\pm$ 30 min)} & \textcolor{lightblue}{MorpheusNet} & \textcolor{lightblue}{0.02}  & \textcolor{lightblue}{79.3} & \textcolor{lightblue}{72.1} & \textcolor{lightblue}{71.6} & \textcolor{lightblue}{94.2}\\
\multirow{4}{*}{} & \textcolor{lightblue}{MorpheusNetQ} & \textcolor{lightblue}{0.02}  & \textcolor{lightblue}{80.6} & \textcolor{lightblue}{74.1} & \textcolor{lightblue}{73.5} & \textcolor{lightblue}{94.6}\\
\multirow{4}{*}{} & XSleepNet2 \cite{xsleep} & 5.6  & 84 & 77.9 & 77.5 & 95.7\\
\multirow{4}{*}{} & SeqSleepNet \cite{xsleep}\cite{{seqsleep}} & 0.2  & 82.6 & 76.4 & - & -\\
\hline


\multirow{4}{*}{Sleep-EDF 78 (in-bed)} & \textcolor{lightblue}{MorpheusNet} & \textcolor{lightblue}{0.02}  & \textcolor{lightblue}{77.8} & \textcolor{lightblue}{71.8} & \textcolor{lightblue}{70.7} & \textcolor{lightblue}{93.8}\\
\multirow{4}{*}{} & \textcolor{lightblue}{MorpheusNetQ} & \textcolor{lightblue}{0.02}  & \textcolor{lightblue}{78.2} & \textcolor{lightblue}{72.7} & \textcolor{lightblue}{71.8} & \textcolor{lightblue}{93.8}\\
\multirow{4}{*}{} & XSleepNet2 \cite{xsleep} & 5.6  & 80.3 & 76.4 & 76.1 & 94.6\\
\multirow{4}{*}{} & DeepSleepNet \cite{xsleep}\cite{eegnet} & 20  & 78.5 & 75.3 & 75.0 & 94.1\\
\hline

\multirow{8}{*}{DOD-H} & \textcolor{lightblue}{MorpheusNet} & \textcolor{lightblue}{0.02}  & \textcolor{lightblue}{83.8} & \textcolor{lightblue}{73.7} & \textcolor{lightblue}{76.5} & \textcolor{lightblue}{95.2}\\
\multirow{6}{*}{} & \textcolor{lightblue}{MorpheusNetQ} & \textcolor{lightblue}{0.02}  & \textcolor{lightblue}{83.8} & \textcolor{lightblue}{75.0} & \textcolor{lightblue}{76.4} & \textcolor{lightblue}{95.2}\\
\multirow{6}{*}{} & SimpleSleepNet (multi-channel)\textsuperscript{\textdagger} \cite{DOD} & 0.094  & 89.9 & 82.4 & 80.2 & 96.4\\
\multirow{6}{*}{} & SimpleSleepNet (single-channel)\textsuperscript{\textdagger} \cite{DOD} & 0.094  & 86.7 & - & - & -\\
\multirow{6}{*}{} & Scorer (average)\textsuperscript{\textdagger} \cite{DOD} & - & 86.5 & 78.56 & - & -\\
\multirow{6}{*}{} & Scorer (best)\textsuperscript{\textdagger} \cite{DOD} & -  & 88.7 & 80.4 & - & -\\
\multirow{6}{*}{} & Scorer (worst)\textsuperscript{\textdagger} \cite{DOD} & -  & 81.7 & 73.7 & - & -
\\\multirow{6}{*}{} & DeepSleepNet\textsuperscript{\textdagger} \cite{eegnet}\cite{DOD}  & 20  & 89.6 & 81.8 & - & -\\
\hline

\multirow{4}{*}{Physionet-2018} & \textcolor{lightblue}{MorpheusNet} & \textcolor{lightblue}{0.02}  & \textcolor{lightblue}{78.6} & \textcolor{lightblue}{76.0} & \textcolor{lightblue}{75.6} & \textcolor{lightblue}{94.0}\\
\multirow{4}{*}{} & \textcolor{lightblue}{MorpheusNetQ} & \textcolor{lightblue}{0.02}  & \textcolor{lightblue}{70.0} & \textcolor{lightblue}{67.7} & \textcolor{lightblue}{67.2} & \textcolor{lightblue}{86.5}\\
\multirow{4}{*}{} & \textcolor{lightblue}{MorpheusNetQ**} & \textcolor{lightblue}{0.02}  & \textcolor{lightblue}{77.3} & \textcolor{lightblue}{74.4} & \textcolor{lightblue}{74.3} & \textcolor{lightblue}{93.7}\\
\multirow{3}{*}{} & XSleepNet2\textsuperscript{\textdagger} \cite{xsleep} & 5.6 & 80.3 & 78.6 & 78.7 & 94.6\\
\hline

\end{tabular}
\end{table*}

\section{Results Summary}
Table \ref{tab:mytable} summarizes the performance in comparison to other models. Models denoted with \textsuperscript{\textdagger} can't be compared directly because of use of different subsets for validation, not using single EEG channel etc.
Overall, our model achieves accuracy values close to more complex models.
As it can be seen, MorphuesNetQ has very close performance to MorphuesNet. However, in Physionet-2018 database, we observed that MorphuesNetQ's performance drop was worse than other databases. We addressed this by not quantizing the start block and the identity block, and this improved performance (indicated by \textsuperscript{**}). 
After the model is trained, it has a size of a about 300KB. This already can be uploaded on more sophisticated hardware platforms. After further compression, Using quantization to 8-bit integers, the final model size was 50KB. With this size, the model can be uploaded onto a range of commercially available MCUs.
To assess embedded deployment, we uploaded the model onto the Arduino Nano 33 BLE development board \cite{arduino}. A digital pin was used to measure the time required to carry out inference. The latency for one inference was 1.6 seconds. 

\section{Discussion}
The model outlined in this paper, MorphuesNet, is a highly compact and optimized model designed for deployment on wearables. It has very few parameters and lower number of arithmetic operations compared to other sleep stage classifiers. As the model is to enable deploying therapies at-home, the model can be fine-tuned to detect certain sleep stages. 

As MorphuesNet has much less parameters compared to other models (shown in Table \ref{tab:mytable}), one can argue that it can't generalize well to databases with larger patient pools. As it is evident in Table \ref{tab:mytable}, MorphuesNet does in fact perform comparable to state of the art. 

At its worst, accuracy drops around 4\%.  MF1, however, shows more drop. This is primarily because of the model's low sensitivity of detecting sleep stage N1. As MorphuesNet only receives the raw EEG signal as its input, it does not receive spectrum based features. We hypothesize that feeding a simple Power Spectral Density (PSD) of the EEG segment would help improve sensitivity. We had empirical findings to confirm this on Sleep-EDF 78 database. 

As the model is designed for carrying out real-time SSC for therapies, there are stages that have been targeted for specific therapy protocols. NREM stages 2 and 3, and REM sleep have been explored for TMR. MorphuesNet has comparable sensitivity to these states (minimum of 0.77 in REM) to state of the art. This indicates that it can be used for sleep-based therapies.

\section{Future work}

This model will be tested on a headset, designed in-house. This headset will be tested against clinical grade PSG setup for comparison of both signal quality, and algorithm performance. This will be a cross-site validation study. 
Further potential improvement would be the addition of EOG as well, as it has shown to improve classification performance in certain cases \cite{xsleep}.

\section{Limitations and Considerations}
It is worth mentioning that when it comes to deployment in an at-home environment, deploying a full PSG setup is problematic. One problem, for instance, is making sure the electrodes are making good contact with the scalp. This is challenging specially when dealing with hair. Clinical PSGs use gel-based wet electrodes. However, in at-home scenarios, EEG collection is restricted to the forehead when using wearables. This implies that the deployed SSC algorithm needs to be trained on classifying sleep stage only using EEG collected from forehead.



\section*{Acknowledgment}
The authors would like to acknowledge the use of the University of Oxford Advanced Research Computing (ARC) facility in carrying out this work.

\end{document}